# Coopetition methodology for resource sharing in distributed OFDM-based cognitive radio networks

Marcin PARZY, Hanna BOGUCKA, *Senior Member, IEEE*

*Abstract*— In this paper, we present a distributed resource allocation mechanism in cognitive radio networks, based on a new *coopetition* methodology, which combines advantages of nodes' competition and cooperation. We postulate that this new method allows for fully distributed resource management between cognitive radio devices. The presented framework is generic, however, we consider it for the application in OFDMA networks. Coopetition takes the best from cooperative and competitive problem formulation, and provides the opportunity to control the balance between fairness and spectral efficiency (SE) of resource allocation. Simulation results confirm that coopetition allows for efficient resource utilization, and may be used practically in wireless cognitive networks.

## I. INTRODUCTION

It is well known that radio resources, such as electromagnetic spectrum are usually limited for practical usage in contemporary radio networks. Therefore it is important to efficiently make use of them. Allocation of these resources should take a number of aspects into account, effectiveness, economy and fairness, being usually most often addressed. Specifically, for the efficient and fair resource allocation, requested Quality of Service (QoS), and links qualities have to be considered, up against the available resources that can be translated to the number of physical channels usable for the applied radio access technologies. Moreover, the network performance metrics are mutually dependent, and reflect contradictory goals of the allocation algorithms: efficient usage of the resources and fairness in their distribution.

This paper considers allocation of resources in the distributed network of Cognitive Radios (CRs), i.e. of intelligent entities, which are able to sense the availability of radio spectrum, and take effective decisions on their usage. In order to acquire information necessary for its efficient operation, a CR node has to have an access to the control (management) channels or to interact and cooperate with other nodes. On the other hand, these control channels and centralized manager may not be available, and the neighboring nodes in the CR network are often competitors in acquiring resources. Thus, the major challenge for a CR node is to operate efficiently while possessing presumably incomplete knowledge on the network environment, and to cooperate with its competitors.

### A. Related work

The centralized spectrum allocation procedures usually apply some optimization procedures that require the Channel State Information (CSI) of all links in the network, and involve a significant amount of the overhead traffic. These procedures are not in our focus, since we concentrate on an uncoordinated CR network, not solely on the opportunistic, but centrally managed spectrum access. Interestingly, game theory provides tools to study competition and cooperation among players taking decisions based on limited or incomplete information [1], [2]. Recent findings show that cooperation among mobile CR nodes in a network can result in their optimized operation, and Pareto-optimal Nash Bargaining Solutions (NBS) or Raiffa-Kalai-Smorodinsky bargaining Solution (RBS) [3] however, exchange of the necessary information requires signaling traffic comparable in volume to the actual information-data traffic. Competitive behavior and distributed decision making can also converge to some desired equilibrium, if the learning process is properly designed, however every learning algorithm requires time to converge, which may not be acceptable in highly-dynamic radio environment. Thus, the game-theory models must be carefully selected and evaluated for the application in the dynamic CR networks.

Let us note that so far a number of game theoretic algorithms have been proposed for spectrum allocation in wireless networks, either cooperative or non-cooperative, however they all assume CSI knowledge of all links or central management or learning in static channels. Our focus is on a decentralized network of nodes that use orthogonal physical channels, such as the ones defined around subcarriers (SCs) in Orthogonal Frequency Division Multiple Access (OFDMA).

A number of non-cooperative, decentralized algorithms for the resource allocation in OFDMA have been considered in the literature [4]-[6]. Usually, the problem is narrowed to centralized SCs allocation, and distributed power allocation based on non-cooperative game models. These models often apply the power-pricing function, which induces fairness among users, or enhances the rate per joule of consumed energy. Alternatively, power allocation problem in OFDMA is defined in such a way, that it can be solved independently by every user and thus, it

Manuscript received June 17th, 2013. This work was supported by the Polish National Centre for Science under the grant agreement no. 779/N-COST/2010/0 for the participation in the European COST IC0902 action.

Marcin PARZY and Hanna BOGUCKA are with the Poznan University of Technology, Poland, (e-mail: mparzy; hbogucka@et.put.poznan.pl).

narrows to an optimization problem.

Some well-known microeconomic oligopoly models can be also applied to address the problems of OFDM spectrum sharing and pricing, i.e. Cournot, Bertrand and Stackelberg game models [7] – [10]. In these models however, the knowledge of the CSI for all users (consisting in channel gains observed by each user at each subcarrier) is assumed, which is not the case in practical systems with distributed decision-making. Study on the game formulation for spectrum sharing and the Nash equilibrium existence is presented in [11]. There, two game models are considered for the distributed spectrum allocation and power loading. One accounts for the partial CSI knowledge (i.e. every user has a perfect CSI of her own channel, and only statistical data on the other users channels) while the other considers only statistical knowledge of the CSI.

Some papers consider distributed dynamic resource allocation in a multi-cell environment, where players are the base stations [12]–[14]. They consider either iterative algorithms for distributed spectrum allocation, power-pricing [13] or the spectrum-usage pricing [14]. The power and spectrum-usage pricing concepts developed for the multi-cell scenario, cannot be considered straight-forward for the single-cell distributed resource allocation, because a base-station has the information of all links in the cell-area, while a mobile node may only have the CSI of its own link.

The major drawback of the solutions proposed for the spectrum allocation in distributed OFDMA networks is that they are found in the non-cooperative games with complete information, and thus assume perfect CSI knowledge of all links among the players, which is not met in mobile scenarios. Some promising decentralized iterative algorithms require some time for the convergence (which must be shorter than the shortest coherence time of all involved links' channels), and in most cases some control traffic to improve or adjust the players' strategies. Note that Bayesian games are also inappropriate because the fading statistics of all channels for all players are required to consider every player behavior with a given probability. In a dynamic radio environment these statistics change with the coherence time. Moreover, it seems impractical to consider the channel gains probability density functions with high granularity because it exponentially increases the complexity of calculating the equilibrium point.

Finally, an important aspect of spectrum allocation is the necessary trade-off between efficiency and fairness of resource distribution. It has been considered last years, e.g. in [15], [16], where the authors addressed this problem by managing the system fairness index.

*B. Paper scope and contribution*

In this paper, new approach to spectrum sharing in distributed OFDMA-based CR networks is proposed. It is based on a methodology called *coopetition*. Coopetition (cooperative competition) is a neologism reflecting a mixture of competition and cooperation. Originally it has been presented in [17]. Coopetition bases on the creation of an added value in cooperation where its distribution is an element of competition. It has been successfully applied in economics, cybernetics, complex production and logistic systems. In [18] we have used this methodology in resource allocation in wireless networks, however reversing the cooperation and competition phases.

In this paper, we extend the work presented in [18] to come out with a generic framework for resource allocation in distributed cognitive radio networks. We focus on OFDMA, and propose to improve the competition phase by applying adaptation of parameters in the Cournot competition. Moreover, we extend the cooperation phase by defining the new coalitional game. Due to the flexibility in the definition of different modes at each stage of coopetition, the considered CR network may operate under different policies, supporting hierarchical traffic, fairness and efficiency of resource utilization.

The paper is organized as follows. In Section II we describe our coopetition framework for radio resource sharing. In Section III, we present a new algorithm used in subsequent coopetition phases for resource sharing in OFDM-based networks. In Section IV, we present the simulation results, and in Section V we conclude our work.

## II. COOPETITION IN DECENTRALIZED NETWORKS

Let us discuss application of coopetition in resource allocation in CR networks. Our players are cognitive devices aiming at acquisition of resources in order to transmit their data with the highest possible data rate. The resources are discretized in the form of a number of the orthogonal frequency channels, e.g. OFDM subcarriers, which are assumed to be flat-fading within their unit bandwidth.

In the proposed coopetition method, the cognitive radio nodes first compete for resources by using non-cooperative complete-information game. The complete information relates to compact metrics representing links qualities. These metrics are required to be representative Channel Quality Indicators (CQI) and not the full CSI representing all channels coefficients. As a result of this competition the players obtain amounts of resources (frequency bandwidth), however, this result is not yet usable, because the players do not know the position of these resources on the frequency axis (frequency carriers). Then, the cooperation phase allows for acquiring the particular carriers by the players. The players cooperate by forming coalitions to tune their spectrum allocations. This method takes advantage of competitive decentralized behavior of nodes with limited signaling but constraint performance, and of the cooperation for improved performance.

The proposed coopetition algorithm consists of four phases: pre-processing (PRP), competition (ComP), cooperation (CooP) and post-processing (PPP). It is assumed that before the algorithm starts, the network nodes have sensed the spectrum resources, and have the information of the available frequency band, as well as on the nature of available orthogonal. The rules of the played games, including the duration of the mentioned phases, are known in the network. These rules can be considered as part of the applied distributed protocol, which may require provision of a common time clock in a network and a beacon (or master) node or a cognitive pilot channel.

In the pre-processing phase, all network nodes calculate the above-mentioned CQI in the available frequency range. These metrics together with parameters reflecting priorities of their data traffic (Traffic Priority Indicators – TPIs) are then exchanged between the nodes, and used in the next phase of the coopetition algorithm. The assumption for these exchanged parameters is that they must be representative and as much compact as it is possible to minimize the signaling overhead. For example, the effective signal to noise ratio (eSNR) can be used as the CQI [19][20]. It represents the channel quality in the whole available frequency band, although it does not provide information on channel selectivity. At the same time this is a very compact metric: a single real-valued number. After PRP, all nodes (players) have complete knowledge on the other players' status, i.e. the CQIs and the TPIs in the network.

After PRP, the competition phase starts. In ComP, non-cooperative game with complete information is played. Here, we have chosen the Cournot oligopoly as the suitable game model based on spectrum pricing [8], [9]. As shown in [21], resource pricing is a remedy for greedy behavior of the players competing for common resources. The pricing as being a part of the game rules are known to the players. Because the complete information relates to the players' CQIs calculated over the whole available band, the result of the Cournot game is the acquired number of orthogonal channels, however, the particular position of these channels on the frequency axis is not determined at this stage. An output of the ComP is the numbers of the players' won assets (frequency channels), which constitute their strengths in the cooperation phase.

The goal of the cooperation phase (CooP) is the transformation of the results of ComP to allocation of the channels on the frequency axis. Here, we propose to apply the coalitional game to decide, how the process of resource allocation is finalized. Note that because the rules of the game are known to the players, and because players know the exact results of the previous phase, they can determine solution of the coalitional game individually, i.e. in a distributed manner (contrary to the approach suggested in [22], where the coalitions of two are formed iteratively by a random choice). The value and strength of a coalition is defined as the number of channels (assets) acquired in the ComP. The stronger the coalition the higher priority in choosing the particular position of their channels on the frequency axis. On the other hand, weaker coalitions have higher flexibility in selecting the particular channels positions, i.e. have more available combinations of allocating these channels. Therefore, the parliamentary coalitions (the weighted-majority) are proposed, allowing for efficient, fair and flexible allocation of channels between coalitions and within a coalition. The result of CooP is the set of coalitions of the CR nodes and a concrete order, in which the coalitions and the coalitionists choose the position of their channels on the frequency axis.

Knowing the strengths of coalitions and the order of spectrum assignments, the players can acquire their channels in the post-processing phase. This can also be done in a distributed manner, because time allocated to sensing of the resources remaining for each coalition, and to the decision on the spectrum allocation for each coalition is determined within the rules of the game. In the PPP, the CR nodes also allocate the power locally to each acquired frequency channel based on the measured frequency characteristic. For this purpose the optimal water-filling algorithm for each single link can be applied.

## III. COOPETITION IN OFDMA

Here, below we present the coopetition algorithm for spectrum sharing in a distributed OFDMA-based uncoordinated cognitive radio network and its mathematical description. Figure 1 presents the algorithm in detail.

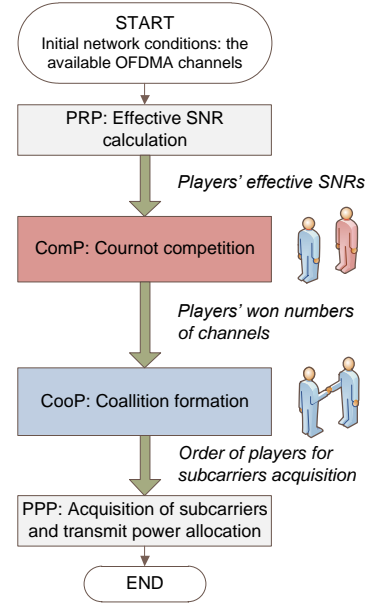

**Figure 1. Coopetition algorithm for spectrum sharing in OFDMA**

### A. The Pre-processing Phase

As an input to the algorithm all nodes must know the available bandwidth $B$ in the sensed frequency range and the number of available SCs, denoted by $K$. The first phase (PRP) starts from the calculation of effective SNRs in the available frequency range, which will serve as the representative CQIs for further coopetition phases. To do this, each node must calculate the optimal power levels for all SCs channels available in the network. The alternative may be calculation of effective SNR (eSNR) based on the constant power for each subcarrier. Each user calculates the eSNR (denoted by $\gamma_{\text{eff},i}$) according to the following formula, [19]:

$$\gamma_{\text{eff},i} = -\beta \cdot \ln\left(\frac{1}{K} \cdot \sum_{k=1}^{K} e^{-\frac{\gamma_{k,i}}{\beta}}\right) \quad (1)$$

where $\beta$ is an adjustment parameter, which depends on the applied modulation and coding scheme, and $\gamma_{k,i}$ is an SNR observed at the $k$th subcarrier of $i$th user. The above definition holds for the exponential – effective SNR mapping method. The eSNR method allows for using single scalar to summarize the whole set of multi-state channel quality measurements. Within this concept, the block error rate (BLER) is measured in a frequency-selective channel with a specific modulation and coding scheme. The eSNR is defined as the SNR that would result in the same BLER, for the said modulation and coding scheme as an AWGN channel. Thus, the eSNR can provide representative knowledge of the channel quality. Note that eSNR defined above serves as CQI for many practical radio resource management algorithms, e.g. it is used in LTE for the basic resource blocks allocations [20].

Moreover, at the considered stage, each player defines her TPI, which in [9] and [10] is called the revenue parameter (denoted by $r_i$) and the target bit-error probability (BEP), denoted by $P_{e,i}$. After calculating the effective SNRs, all nodes broadcast the triples ($\gamma_{\text{eff},i}$, $r_i$, $P_{e,i}$) to the other nodes. This triplet is right the reduced information made known to all players, and allows for playing the complete-information non-cooperative game between the CR nodes in the next phase.

### B. Competition Phase

In the competition phase, the Cournot game is played. Definition and analysis of this game is presented in [23], [24]. It has been adopted to radio resource-sharing in [8] – [10]. Here below, we extend work presented in [8]. In the proposed game model, the players are the network nodes, their strategies consist in the number of acquired SCs, and their payoff functions reflect their profit, i.e. revenue minus cost, where the revenue relates to the obtained transmission throughput, and cost relates to the pricing of acquired spectrum resources.

In the first step, of the Cournot competition every player estimates the spectral efficiency (SE) $\eta_i$ based on ($\gamma_{\text{eff},i}$, $r_i$, $P_{e,i}$) triplets for every player $i$ [25]:

$$\eta_i = \log_2(1 + \alpha_i \cdot \gamma_{\text{eff},i}), \quad (2)$$

where for QAM modulation schemes:

$$\alpha_i = \frac{1.5}{\ln\left(\frac{0.2}{P_{e,i}}\right)}. \quad (3)$$

After calculating the spectral efficiencies, each player analyzes the profit function (denoted by $\pi_i$) expressed by:

$$\pi_i(\mathbf{b}) = r_i \cdot \eta_i \cdot b_i - c(\mathbf{b}), \quad (4)$$

where $b_i$ is the amount of spectrum acquired by the $i$th player (this is player $i$ strategy, and an unknown variable), $c(\mathbf{b})$ is the cost function dependent on the set of strategies $\mathbf{b} = \{b_1, \ldots, b_i, \ldots, b_I\}$, and $I$ is the number of players. The cost function is given by the following formula:

$$c(\mathbf{b}) = x + y \cdot \left(\sum_j b_j\right)^\tau, \quad (5)$$

where $x$ reflects the fixed cost of spectrum sharing (expressed in some monetary units), $y$ is the cost of spectrum unit, and $\tau$ is the factor, which impacts the cost of spectrum sharing. Note that these pricing parameters ($x$, $y$, $\tau$) are assumed to be known, by the players, since they are part of the network protocol. They can be linked to the physical quantities of the network, such as the available spectrum bandwidth $B$ the number of players $I$ and their effective SNRs $\gamma_{\text{eff},i}$. This is because in the considered economical model, $B$ is the *supply* quantity in the market, while $I$ and $\gamma_{\text{eff},i}$ impacts the spectrum *demand* $b_i$ ($\gamma_{\text{eff},i}$ reflects usefulness of the spectrum to player $i$). The smaller $B$, the higher pricing-parameters values $x$ and $y$, what is represented by the market *inverse demand function* [9]. Moreover, for a given $B$, higher demands $b_i$ must be counterbalanced by higher $x$ and $y$.[1] Every player aims at maximizing her profit as a response to given strategy choice of the other players, i.e. her optimization task is defined as follows:

$$\text{find } b_i^* \text{ s.t. } \pi_i(\mathbf{b}_{-i} \cup b_i^*) = \max_{b_i} \pi_i(\mathbf{b}), \quad (6)$$

where $\mathbf{b}_{-i}$ is the set of strategies of all other players excluding player $i$. The profit $\pi_i$ is a concave function of $b_i$ (for $\tau = 1$), and thus, the maximum can be found by solving equation:

$$\frac{\partial \pi_i}{\partial b_i} = 0. \quad (7)$$

The best response function for the $i$th player (the function indicating the best payoff strategy $b_i$ given the set of strategies chosen by all other players) for $\tau = 1$ is given by:

$$b_i^* = \frac{1}{2} \cdot \left(\frac{r_i \cdot \eta_i - x}{y} - \sum_{i \neq j} b_j\right). \quad (8)$$

Equation (8) for $i = 1,\ldots, I$ forms a set of $I$ equations with $I$ unknown variables:

$$\mathbf{A} \cdot \mathbf{b}^{*\mathrm{T}} = \mathbf{z}, \quad (9)$$

where $\mathbf{b}^{*\mathrm{T}} = (b_1^*,\ldots, b_i^*,\ldots,b_I^*)^{\mathrm{T}}$, $\mathbf{z}$ is a vector of elements given by the following formula:

$$z_i = \frac{r_i \cdot \eta_i - x}{y}, \quad (10)$$

and $\mathbf{A}$ is a square matrix ($I \times I$) with elements given by:

$$a_{i,j} = \begin{cases} 2 \text{ for } i = j \\ 1 \text{ otherwise.} \end{cases} \quad (11)$$

Because matrix $\mathbf{A}$ is a rank one modification of the unity matrix, the elements of vector $\mathbf{b}^{*\mathrm{T}}$ can be obtained as:

$$b_i^* = z_i - \frac{\sum_i z_i}{1+I}. \quad (12)$$

This solution is unique, and can be obtained by every player. It is also the Nash Equilibrium (NE) for the considered game.

For $\tau \neq 1$ the numerical methods must be used, which are more computationally expensive. In case if there is only one player (monopoly case) the following solution can be applied (where there is no need for limiting the solution only to $\tau = 1$).

$$b_i^* = \sqrt[\tau]{\frac{r_i \cdot \eta_i - x}{y \cdot (\tau+1)}} \quad \text{for} \quad \frac{r_i \cdot \eta_i - x}{y \cdot (\tau+1)} \geq 0. \quad (13)$$

In searching for the $b_i^*$ value, the following constraints must be considered. The value $b_i^*$ must be nonnegative, and cannot exceed $B$. Moreover, the spectrum assignments must be multiples $k_i$ of an OFDM subchannel bandwidth $\Delta f$. Thus,

$$\forall_i \ 0 \leq b_i^* \leq B \quad (14)$$

$$b_i^* = k_i \Delta f. \quad (15)$$

Here, integer-programming methods ($k_i$ is a positive integer) can be applied. Alternatively they can be handled by the two following simplified approaches.

In the first approach, the results of the unconstraint Cournot game ($b_i^*$) are rescaled by the factor being the fraction of the number of available spectrum units $K$ (orthogonal channels or SCs) over the sum of all players' demands $D$ resulting from this game. Then, an integer part of the calculated spectrum units is reserved for each player, since the number of SCs $s_i$ assigned to player $i$ has to be non-negative integer. This rescaling is particularly useful in cases when not all resources were acquired as a result of the unconstraint Cournot competition or when these results exceed the available bandwidth, $D > B$ (pricing

---

[1] Typical values of the pricing parameters $x$ and $y$ for some supply-demand scenarios in a CR network are given in [9].

TCOM-TPS-13-0451.R3                                                                                          5

parameters values are too low or revenue parameters are too high). Rescaling causes that all (or almost all) the resources are allocated proportionally to the results of the unconstrained Cournot competition. The complexity of this algorithm is very low because of the simplicity of the formulas presented in equations (12) and (13). The algorithm described above is presented in Table I, and will be called the Cournot game classic solution scaling. Note that the complexity of finding the NE for the classical Cournot competition increases linearly with the number of players $I$. This also holds for the Cournot Game Classic Solution Scaling.

The second approach to meet constraints (14)-(15) is based on adaptive pricing parameters ($x$ and $y$) adjustment to obtain the spectrum assignments such that minor bandwidth is left unoccupied. The respective algorithm is presented in Table II.

TABLE I
COURNOT GAME CLASSIC SOLUTION SCALING

| Step | Step description |
|---|---|
| 1. | if $b_i^* \leq 0$ then $b_i^* := 0$ |
| 2. | if $b_i^* \geq B$ then $b_i^* := B$ |
| 3. | $D := \sum_i b_i^*$ |
| 4. | $\forall_i\ s_i := \left\lfloor \frac{K \cdot b_i^*}{D} \right\rfloor$ |

TABLE II
COURNOT GAME PARAMETERS ADAPTATION

| Step | Step description |
|---|---|
| 1. | Set $x := 0$, $y := 1$ |
| 2. | Find the NE strategies in the Cournot game: $b_i^*$ and $D := \sum_i b_i^*$ |
| 3. | while ($D > B$) repeat<br>$y := y + \Delta y$<br>Find the NE strategies: $b_i^*$ and $D$ |
| 4. | $x_1 := x$, $y_1 := y$, $D_1 := D$ |
| 5. | if ($y > 1$)<br>$y := y - \Delta y$<br>Find the NE strategies: $b_i^*$ and $D$ |
| 6. | while ($D > B$) repeat<br>$x := x + \Delta x$<br>Find the NE strategies: $b_i^*$ and $D$ |
| 7. | $x_2 := x$, $y_2 := y$, $D_2 := D$ |
| 8. | if ($D_1 \geq D_2$)<br>$x = x_1$, $y = y_1$<br>else<br>$x := x_2$, $y := y_2$ |
| 9. | Find the NE strategies: $b_i^*$ and $D$ |
| 10. | $\forall_i\ s_i := \left\lfloor \frac{K \cdot b_i^*}{D} \right\rfloor$ |

In the first line, the pricing parameters are initiated. Then, the NE is found, as described above. If the sum of the assigned bandwidth is higher than $B$ the pricing parameter $y$ is incrementally increased by $\Delta y$ until this condition is satisfied. If $y$ is higher than 1 its value is decreased. The pricing parameter $x$ is then similarly optimized using incremental value $\Delta x$ until the sum of assigned resources does not exceed $B$. The sum of bandwidth assignments $D$ resulting from optimization of $y$ and of $x$ are compared, and the solution which is closer to $B$ is chosen. Finally the NE is found for the Cournot game using this set of parameters and the number of SCs assigned to each player is calculated. Note that initial values of $x$ and $y$, as well as $\Delta x$ and $\Delta y$ are parameters of the CR network protocol, and are known in the network, e.g. through a cognitive pilot channel or master node broadcast signaling.

The algorithm described above allows for adaptive adjustment of parameters in the Cournot model in order to obtain assignments of SCs not exceeding the total number of available SCs. The iterative algorithm from Table II invokes the algorithm in Table I in each iteration. Thus, its complexity is approximately $\omega$ times higher, where $\omega$ is the number of iterations, and it is linearly scalable with the number of players $I$.

Iterative adjustment of $x$ and $y$ in the Cournot Game Parameters Adaptation algorithm presented in Table II optimizes the spectrum usage. One can envision optimization of these parameters to find the maximum of the goal function in the network, e.g. sum-throughput or the SE. Solution of this problem would mathematically define dependence of $x$ and $y$ on network physical quantities such as the available bandwidth, number of players and their eSNRs. This problem is however very complex, and in general does not have a closed-form solution. In [26], a related problem of the optimization of $y$ to maximize the above-mentioned goal functions is defined, where $x = 0$ and $\tau = 1$. The closed-form solution could only be found for a two-path channel model. Thus, numerical optimization of these parameters is necessary. Moreover, some averaging needs to be done to find good (nearly-optimal) $x$ and $y$ values for a number of channel- and user types, so that these values could be precalculated and stored to avoid extensive computations each time the channel conditions and the user-types change.

Let us now propose the variants of the Cournot game which increase fairness in resources allocation. The first variant is an oligopoly competition described above. It will be called Cournot type 1 in the remainder of this paper, and, as shown above, can be considered in the classic form (denoted by C1c and defined by the algorithm in Table I) or in an adaptive one (denoted by C1a, and defined in Table II). There is no fairness mechanism introduced in the Cournot type 1 oligopoly case.

In the second variant, we propose to introduce the minimum spectrum assignments for each player. The minimum number of SCs is assigned to each user when $b_i^*$ is lower than a considered fraction of SCs resulting from (13). The game with minimum SCs assignments will be called Cournot type 2 in the remainder of this paper, and again it may be considered in the classic (C2c) version with no adaptation of the game parameters or in the adaptive version (C2a).

In order to further increase the fairness, we propose to introduce and examine the so-called monopoly case. In this case, the profit function for a single player depends only on this player's demand and her pricing parameters values. The values $b_i^*$ are calculated according to (13). This game for $\tau = 1$, $\tau = 2$, $\tau = 3$ will be called Cournot type 3, Cournot type 4 and Cournot type 5 respectively. All these games may be in the classic or in adaptive version. Increase of the $\tau$ value causes the increase of fairness, but decreases the SE.

To summarize this section, in Cournot 1, the players with high revenue parameters (high TPI) and high effective SNR are prioritized. The other games from Cournot 2 to 5 assure better fairness. The final result of the ComP is the number of resource units (SCs) assigned to each user – $s_i$.

*C. Cooperation Phase*

In the design of the cooperation phase, the main goal is to determine the order of accessing the spectrum resources, which leads to the final SCs allocation. The first option is to rank the players according to the number of SCs acquired in the ComP in the descending order (from the one with the highest $s_i$ – called "the strongest" to the one with the lowest $s_i$ – called "the weakest"). The alternative to this is the ascending order of the players' "strength".

Let us now consider the coalition-building method. An interesting and promising option is the weighted-majority game [1] to form coalitions that can obtain the majority of resources, yet having enough flexibility in sharing them between the coalitionists. Here, $s_i$ values are used as weights of the players. These values represent the *strengths* of the players, and, as discussed above, the number of OFDM SCs acquired in the ComP. It is assumed that the winning coalition will have the right to take decision before the other players on the particular SCs to be occupied by this coalition. After finding the winning coalition, the players of this coalition are excluded from further consideration (their priority in choosing the SCs is already defined), and the next round of the coalition formation takes place to find the next-priority set of players. The indices of the subsequent stages of coalition formation can be denoted by $n = 1, ..., N$, and possible-coalitions indices at each stage are denoted by $m = 1, ..., M^{(n)}$. Note that at stage $n$, there are $I^{(n)}$ users and $K^{(n)}$ subcarriers ($I^{(1)} = I$, $K^{(1)} = K$).

Let us now define the coalition "strength". The $m$th coalition strength at the $n$th stage $B_m^{(n)}$ is measured in the number of the assets of its participants, equal to the number of spectrum units the coalition is entitled to choose:

$$B_m^{(n)} = \sum_{i \in \mathbf{S}_m^{(n)}} s_i , \quad (16)$$

where $\mathbf{S}_m^{(n)}$ is the set of players in the $m$-th coalition at stage $n$.

Apart from the coalition strength, let us define the coalition *flexibility*. The $i$th player flexibility (of choosing her spectrum) in coalition $m$ is expressed as the number of combinations she has at the $n$th turn, when there are $K^{(n)}$ spectrum units left for the coalition, and $K_{m,i}^{(n)}$ is left for player $i$ in coalition $m$:

$$f_{m,i}^{(n)} = \binom{K_{m,i}^{(n)}}{s_i} = \frac{K_{m,i}^{(n)}!}{s_i! \left( K_{m,i}^{(n)} - s_i \right)!} . \quad (17)$$

For example, at the beginning of the coalition formation, in the winning coalition, the first player will have $K$ available SCs, the second one in this coalition will have $K$ minus the resources taken by the first player, etc. For a player it is also important to calculate her order (priority) in choosing the spectrum units when she joins the coalition. Let us note that the order, in which player $i$ chooses her spectrum, when she joins coalition $m$ at stage $n$ (denoted as $p_{m,i}^{(n)}$) can be determined either according to her strength or to her flexibility.

Finally, the flexibility of the $m$th coalition is defined as the number of combinations the coalition has at the $n$-th turn, when there is $K^{(n)}$ spectrum units left for its acquisition:

$$F_m^{(n)} = \binom{K^{(n)}}{B_m^{(n)}} = \frac{K^{(n)}!}{B_m^{(n)}! \left( K^{(n)} - B_m^{(n)} \right)!} . \quad (18)$$

Let us now consider the players' reasoning when forming coalitions. This reasoning can be summarized as follows:

- Every player would like to be in the coalition that would guarantee her higher priority (lower order) $p_{m,i}^{(n)}$ in choosing the spectrum units than that in all other cases,
- Every player would like to have the highest possible flexibility in choosing her spectrum units,
- Every player must benefit from forming a particular coalition, otherwise she would not form it.

As mentioned above, our game is the weighted-majority game, and thus, at each stage, the value of a coalition equals 1 if it is the winning coalition and 0 if it is not. A coalition of players is winning if the sum of the weights of its members (coalition strength) exceeds the predefined positive quota $\mu < 1$, which for the regular majority equals 0.5. Note that there may be multiple coalitions for which the coalition strength exceeds the quota, and for which the conditions of joining the coalition (listed above) are satisfied. Because apart from the number of SCs to be selected, flexibility in this selection is very important for the coalitionists, we postulate that the winning coalition must also have the highest flexibility out of all possible majority coalitions. Thus, at the $n$th stage, the $m$th coalition $\mathbf{S}_m^{(n)}$ value is defined as follows:

$$v\left(\mathbf{S}_m^{(n)}\right) = \begin{cases} 1 & \text{if } \frac{B_m^{(n)}}{B^{(n)}} > \mu \text{ and } \forall_{j \neq m} \frac{B_j^{(n)}}{B^{(n)}} > \mu \Rightarrow F_m^{(n)} \geq F_j^{(n)} \\ 0 & \text{otherwise} \end{cases}, \quad (19)$$

where $B^{(n)} = \sum_{i \in \mathbf{S}^{(n)}} s_i$ (with omitted index $m$) is the sum of weights (strengths) of all players' playing the weighted-majority game at stage $n$, and $\mathbf{S}^{(n)}$ is the set of these players. Thus, the coalition formation in weighted-majority game at stage $n$ is the multidimensional optimization problem, i.e. for $j = 1, ..., M^{(n)}$:

$$\begin{aligned} \text{find } \mathbf{S}_m^{(n)} \text{ s.t. } F_m^{(n)} &= \max_j F_j^{(n)} \\ \text{subject to: } \frac{B_m^{(n)}}{B^{(n)}} &> \mu . \end{aligned} \quad (20)$$

Note that for $\mu = 0.5$ the coalition that has the lowest $B_m^{(n)}$ exceeding quota $0.5 B^{(n)}$ will have the highest flexibility $F_m^{(n)}$ out of the coalitions exceeding the quota, i.e.

$$\begin{aligned} \forall_{m,j}: \frac{B_m^{(n)}}{B^{(n)}} &> 0.5, \quad \frac{B_j^{(n)}}{B^{(n)}} > 0.5 \\ B_m^{(n)} &< B_j^{(n)} \xrightarrow{\text{yields}} F_m^{(n)} > F_j^{(n)} . \end{aligned} \quad (21)$$

Thus, in this case, at the $n$th stage of the coalition formation, the strength of the winning coalition should exceed $0.5 B^{(n)}$ by minimum possible value to assure its high flexibility. This will give the winning coalition a high number of possible choices in selecting favorable SCs. The rule should be to consider all possible coalitions for each stage and to find the winning one as defined by (19). In general, the higher the value of $\mu$, the higher the strength of the winning coalition at a single stage, the more players are usually involved in it, and thus, the lower number of stages in the coalition formation process. Note that there are methods to simplify the coalition formation process, e.g. the merge and split algorithm, described in [27].

Finally, it is important to note that although for clarity of considerations we have introduced coalition formation in stages, all players can implement the coalition formation algorithm at the same time, without waiting for the results from other players or stages. This is because all players have exactly the same data resulting from ComP. The algorithm of coalition formation is presented in Table III.

**TABLE III**
**COALITION FORMATION ALGORITHM**

| Step | Step description |
|---|---|
| 1. | Consider all possible coalitions of the players, who are not yet part of any coalition satisfying $\frac{B_m^{(n)}}{B^{(n)}} > \mu$; |
| 2. | Rescind coalitions containing players, who have not improved their priority $p_{m,i}^{(n)}$ or flexibility $f_{m,i}^{(n)}$ by joining a coalition; |
| 3. | Calculate $F_m^{(n)}$ of the majority coalitions and choose the one with the highest $F_m^{(n)}$; |
| 4. | If there are more than one prevailing coalitions, for each coalition consider average eSNR and choose coalition with the highest average eSNR. |
| 5. | Form the strongest coalition with the highest flexibility, and determine the order of acquiring SCs:<br>– from the strongest player in a descending order of players' strengths (if two players have the same strength start from the player with higher eSNR), or alternatively<br>– from the player with the highest flexibility in the descending order of the players' flexibility; |
| 6. | Tag the players of the formed coalition as the ones already involved. Go to Step 1. |

The result of this cooperation phase allows for the determination of the order of players to acquire SCs. The main goal for this phase is to mix the players with the high number of resources acquired in the competition phase (i.e. in average having higher eSNR than others and higher TPI) with the players having worse channel conditions and lower TPI. It improves fairness in resource sharing. If weak players belong to the strong coalition, they are allowed to acquire the same (small) number of SCs (due to the *non-interchangeable assets* rule applied), however, they have higher chance to choose their best SCs before these SCs are occupied by other players. Thus, although the Cournot competition assures more resources to the players of the higher eSNR and having higher revenue parameter values, the fairness of spectrum allocation can be improved at the cooperation stage by letting weak players form a coalition with the strong ones.

### D. Post-processing Phase

In the post-processing phase, the players can acquire their SCs. Note that after ComP, all players know each other's strengths, and thus can locally analyze coalition formation in CooP. Consequently, each player in the PPP knows her turn in acquisition of SCs. Thus, this final SCs allocation can be done individually in a distributed manner provided that there is a common clock signal in the network to synchronize subsequent turns.

In the PPP, the power is also allocated to players and their SCs. The well-known optimized solution is the multiuser water-filling [28], however it is centralized and computationally complex, and thus, not in our focus. A simple distributed method is to assign the players the same power irrespectively of the number of their SCs. This way, the players with lower number of SCs can improve their throughputs. A more efficient option is to allocate transmit power to players proportionally to the number of acquired assets. This solution favours the players having higher channel quality. In this approach, each user allocates the assigned power locally according to the water-filling principle.

To conclude our considerations in this section, let us note that despite some optimization mechanisms implemented in the competition and cooperation phases described by (6) and (20) respectively, it is not enough to call the whole process (the sum of competition and cooperation, namely *coopetition* methodology) optimal. The proposed methodology can be called *heuristic* and practical algorithm. This is because the main assumption for the design of the coopetition algorithm is to limit the exchange of necessary information and associated signalling, and still to improve efficiency and fairness of resource sharing. Note that only the compact metrics are transmitted before the competition phase, i.e. CQI (effective SNRs), TPI and the target BEP. This allows the nodes to take decisions individually. The distributed nature of the Cournot competition does not allow to use its result efficiently, because it consist in the number of SCs acquired by the players, but not their particular positions on the frequency axis. However, the knowledge of this unique result by every player allows them to further cooperate for the efficiency of spectrum sharing, i.e. to acquire particular favourable OFDM SCs. To this end, the players analyse formation of the majority coalitions in a distributed way, i.e. no exchange of the information between the players is needed, and every player can implement the coalition formation algorithm locally with the same unique result. Therefore, the proposed coopetition framework can be called distributed. Moreover, it provides means for balancing SE in the network and fairness.

### IV. SIMULATION RESULTS

#### A. Simulation setup

We have verified the coopetition method in an OFDM-based CR network through Monte-Carlo computer simulations assuming the following network parameters. An LTE-like system has been considered with 15 kHz of subcarrier spacing, and *K* = 300 available SCs. The LTE Extended Vehicular A channel model defined in [29] has been chosen with the delay profile of [0, 30, 150, 310, 370, 710, 1090, 1730, 2510] ns, and related power profile of [0, –1.5, –1.4, –3.6, –0.6, –9.1, –7, –12, –16.9] dB, Doppler frequency = 0 Hz. The parameter $\beta = 30$ has been adopted for the eSNR calculation.

The following parameters and assumptions have been made for the Cournot oligopoly games. For Cournot game types 1–5 described in the previous section, the pricing parameters $x = 0$, $y = 1$ have been adopted in case of the Cournot game classic solution scaling. The same values have been adopted as initial pricing parameters in the Cournot Game Parameters Adaptation, while $\Delta x = \Delta y = 0.5$ have been chosen, as resulting from the assumed number of iterations in this algorithm ($\omega$ equal to 5 for 8 players). As mentioned before, for the Cournot game type 1 and 2 $\tau = 1$ has been assumed, while for the Cournot game type 3, 4 and 5, $\tau = 1$, $\tau = 2$ and $\tau = 3$ has been adopted respectively. The target BEP equal to $10^{-4}$ has been assumed for all links. Various revenue parameters have been assumed for the CR network nodes depending on the chosen scenarios defined in the next subsection. The average SNRs have been randomly chosen and have resulted in random effective SNRs in the range between 5 and 25 dB.

The coalition formation has been modeled as described in the previous section. In the final post-processing phase, the power allocation to the players has been assumed to be based on the proportional assignment (proportionally to the fraction of acquired SCs), and the local water-filling.

Our coopetition algorithm has been compared with the Round-Robin (RR), MaxSNR, the cooperative algorithm based on NBS as described in [3], the non-cooperative Cournot competition without coalition formation (with either *weakest-player-first* or *strongest-player-first* rule for the priority in acquiring resources), and with the cooperative algorithm based on the Shapley value [30] for resource distribution as described in [31]. The value of coalition considered in [31] reflects the benefit of the players in terms of the transmission sum-rate achievable for the coalition with respect to the claims in the network. For our simulations, it has been assumed that the claims are rates that can be achieved ("claimed") by the players given their CQIs, TPIs, target BEPs (($\gamma_{\text{eff},i}$, $r_i$, $P_{\text{e},i}$) triplets) and equal share in the available bandwidth expected ("claimed") by the players. Acquisition of SCs is then done based on the *strongest-player-first* rule. Although the first three mentioned algorithms are the centralized ones requiring complete knowledge of the players' channels characteristics, they can serve as reference algorithms providing good fairness (Round-Robin), high SE (Max SNR) or a cooperative solution (NBS). For the assessment of fairness, Jain's index has been calculated, as defined in [32].

The following network scenarios have been considered:
- Scenario 1: $I = 8$ players with no priorities (equal revenue parameter $r_i = 1$) and the average SNR (averaged over the available SCs) equal to 15 dB;
- Scenario 2: $I = 8$ players with no priorities, where 4 players observe the average SNR of 10 dB and 4 players – of 20 dB;
- Scenario 3: $I = 8$ players with no priorities, with a random average SNR uniformly distributed between 5 and 26 dB;
- Scenario 4: $I = 12$ players with no priorities, with a random average SNR uniformly distributed between 5 and 26 dB.
- Scenario 5: $I = 8$ players with various TPIs (every 2 players are characterized by $r_1 = 0.8$, $r_2 = 1.6$ $r_3 = 2.4$ and $r_4 = 3.2$ respectively) and random average SNRs uniformly distributed between 5 and 26 dB.

Finally, as said before, for each scenario the Cournot competition has been implemented in the following modes: C1c, C1a, C2c, C2a, C3c, C3a, C4c, C4a, C5c, and C5a.

### B. Simulation results

In Scenario 1, the SE and fairness is similar for all coopetition modes (see Fig. 2 – 3). Moreover, the resulting SE of the coopetition approaches is similar to cooperative NBS, and slightly lower than the SE of cooperative algorithm using the Shapley value for the resource distribution.

When 50% of players have better channel conditions than the others as in Scenario 2, the cooperative NBS results in the lowest SE and the highest fairness, while for the cooperative algorithm based on Shapley value both SE and fairness are medium. The coopetition approach balances between fairness and efficiency (Fig. 4 – 5).

In Scenarios 3 and 4, the channel qualities are random resulting in disparities between the different numbers of players. In these scenarios presented in Fig. 6 – 9, the SE of the coopetition approach can be quite close to that of MaxSNR algorithm, while its fairness can be close to the Round-Robin algorithm for some Cournot games types. When compared to the cooperative algorithm based on Shapley value, coopetition using some Cournot games types may achieve higher SE and lower fairness, while for the other types it is the opposite. In Scenario 5, the players' TPIs have impact on the achieved SE and fairness (Fig. 10 – 11) which is slightly lower than in Scenario 4 for the same number of players.

Finally, note that in the resource sharing options there is always a contradiction between SE and fairness, what is shown in Fig. 12 – 13 for example Scenario 3 and 5. There, the sets of the same three markers relate to the same Cournot game type. In each set, the point with the highest SE and the one with the highest fairness relate to the results of the competitive behavior with the *strongest-first* rule in spectrum acquisition and the *weakest-first* rule respectively. The point in between relates to the results of coopetition with coalitions.

Note that in all scenarios, the mode of cooperation phase allows for better control between fairness and efficiency, and a proper mode selection is a matter of applied policy in a network. Despite of the distributed nature of our approach, efficiency and fairness can be well balanced. We believe that this is the major achievement and contribution of the paper. The centralized algorithms which either maximize SE, and completely ignore fairness (MaxSNR), or contrary handle fairness, and completely ignore efficiency (Round Robin) cannot guarantee such a result.

### C. Complexity and convergence assessment

The complexity in terms of the number of real operations (either multiplications or additions, which require one clock cycle in a Digital Signal Processor – DSP) depends on the chosen Cournot game type and its parameters, as well as on the network scenario parameters: $I, K, \mu, \tau$. In the worst case, when the players have equal CQIs, TPIs and target BEPs the number of steps $N$ in the coalition formation is the highest, and the total number of real operations $\Lambda_{tot}$ in the considered coopetition methodology is approximated as follows:

$$\Lambda_{tot} \approx \Lambda_C + \sum_{n=1}^{N} \sum_{l=1}^{L^{(n)}} \left[ 2 \binom{L^{(n)}}{l} l + 1 \right] + \\ + \sum_{n=1}^{N} \left\{ \binom{L^{(n)}}{\lceil \mu L^{(n)} \rceil} [1 + (\mu K)^2 (1-\mu)^{2n}] \right\}, \quad (22)$$

where $\Lambda_C$ is the number of operations of the Cournot game, which for $\tau = 1$ and the Cournot Game Classic Solution Scaling equals $(3I + 5)I$, and for the Cournot Game Parameters Adaptation equals $(3I + 2)\omega I + 3I$, and $\omega$ is the number of iterations in the adaptive algorithm, which can be adjusted depending on $K$, $I$ and the assumed accuracy. When $\tau > 1$ additional operations are required to compute the root in (13). If the Newton-Raphson algorithm is used for this purpose, its

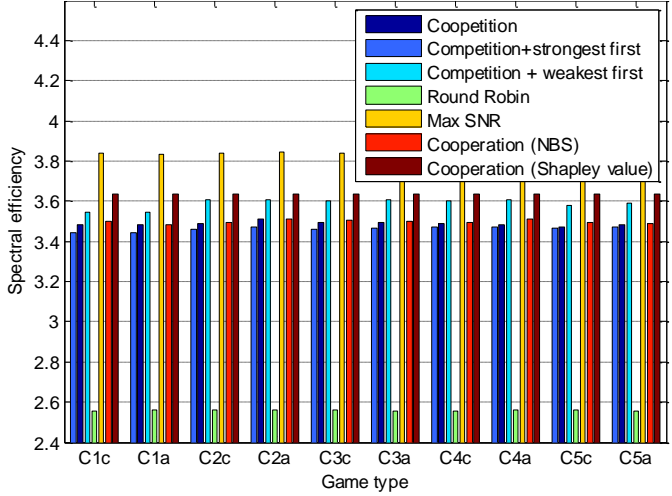

**Figure 2. Spectral efficiency vs. Cournot game type in coopetition Scenario 1.**

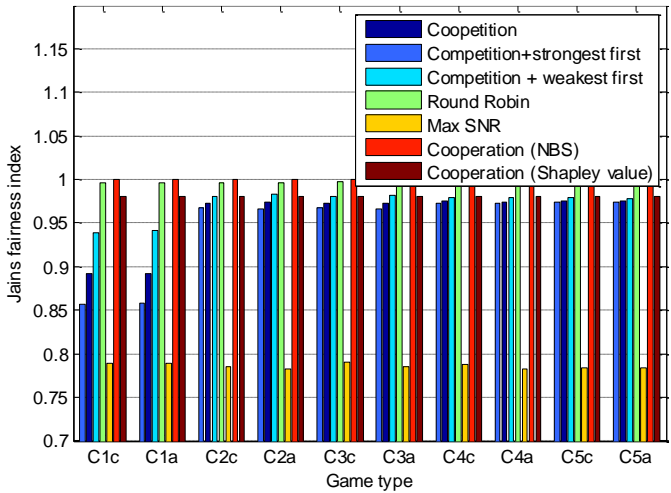

**Figure 3. Jain's fairness index vs. Cournot game type in coopetition Scenario 1.**

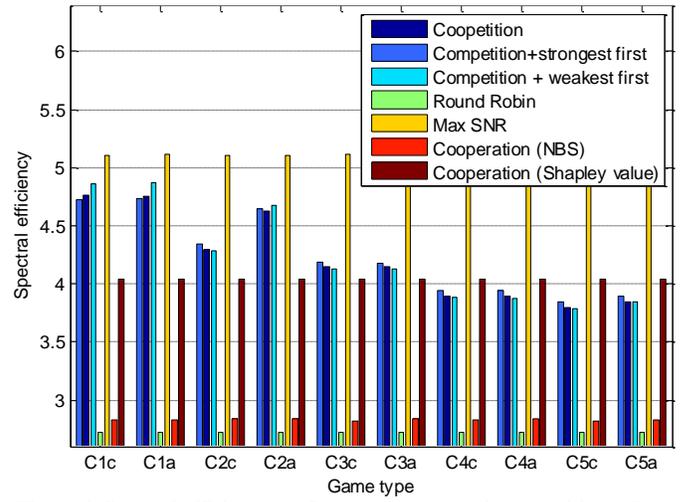

**Figure 4. Spectral efficiency vs. Cournot game type in coopetition Scenario 2.**

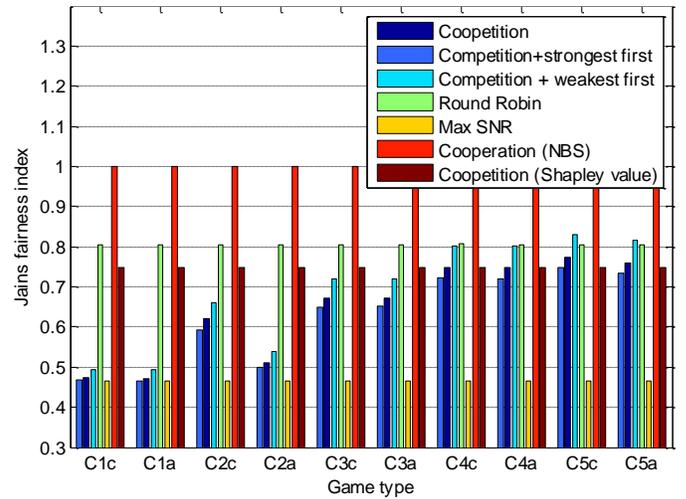

**Figure 5. Jain's fairness index vs. Cournot game type in coopetition Scenario 2.**

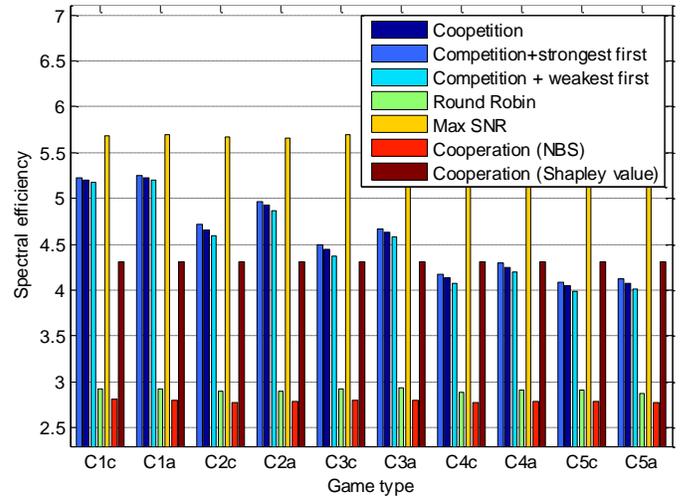

**Figure 6. Spectral efficiency vs. Cournot game type in coopetition Scenario 3.**

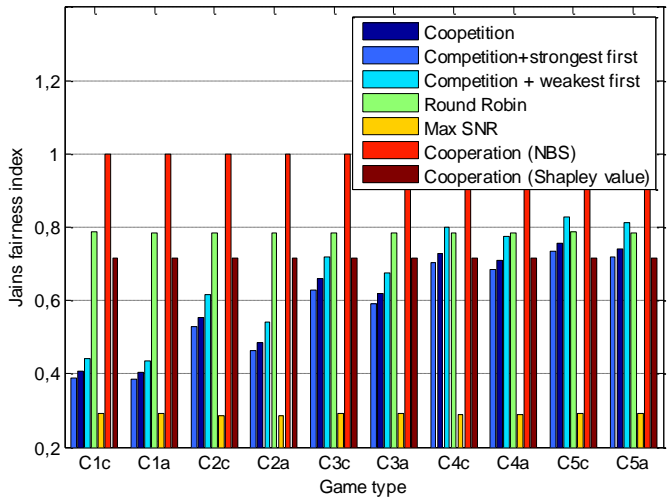

**Figure 7. Jain's fairness index vs. Cournot game type in coopetition Scenario 3.**

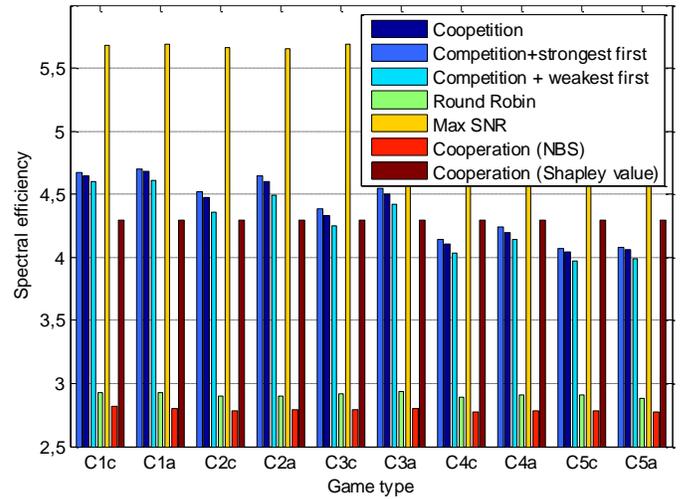

**Figure 10. Spectral efficiency vs. Cournot game type in coopetition Scenario 5**

each iteration requires $\ln(\tau)$ multiplications and one division. Moreover, in formula (22), $L^{(n)} = (1-\mu)^n I$ is the expected number of players participating in coalition formation at stage $n$, $N = -\log_{1-\mu} I$ is the expected number of steps in the coali-tion formation, and ⌈·⌉ denotes the *ceiling* function. Note that $L^{(n)}$ is geometric progression with a common ratio $1 - \mu < 1$. Thus, even in the worst case discussed above, the coalition formation process converges (to a single player at stage $N$) at the geometric speed. The second component in (22) reflects the number of operations necessary for the calculation of strengths of all possible coalitions $B_m^{(n)}$ (including the grand coalition strength $B^{(n)}$) at all stages in the CooP, as well as the number of operations necessary for the calculation of the $B_m^{(n)} : B^{(n)}$ fractions and for the comparison of these fractions with $\mu$, i.e. the complexity of the first step in Table III. The third component in (22) reflects the number of operations necessary for the calculation of flexibilities of all majority coalitions at all stages and for comparison of the coalitions flexibilities (steps $2-3$ in Table III). Note that complexities of steps $4-6$ are minor in comparison with steps $1-3$ in that table. Coopetition can be successfully implemented in a mobile CR node if $\Lambda_{tot} T_{clock} \leq T_c$, where $T_{clock}$ is the period of the DSP clock cycle, $T_c$ is the channel coherence time. In case of the parameters used in our simulations (given in Section IV.A), the total worst-case number of operations $\Lambda_{tot}$ can be estimated as not exceeding approximately 38 000 irrespectively of the Cournot game type, since the major complexity is associated with the coalitional game. Thus, assuming example DSP clock frequency of 1.2 GHz (e.g. in power-optimized TMS 320C6000), the algorithm in our simulated network can be implemented for channels, where $T_c \geq 31.54$ μs.

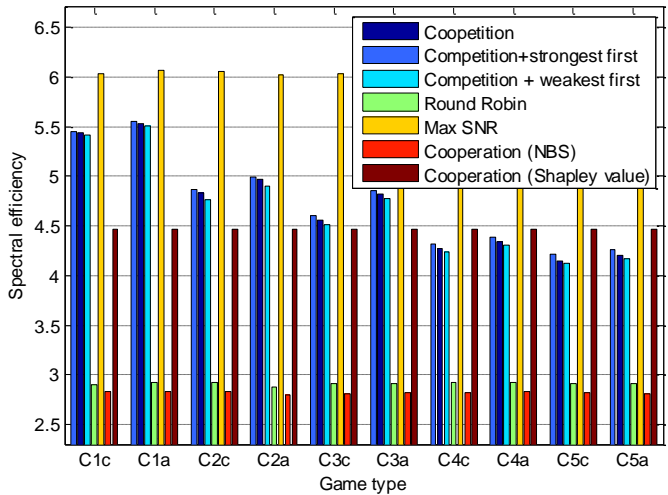

**Figure 8. Spectral efficiency vs. Cournot game type in coopetition Scenario 4.**

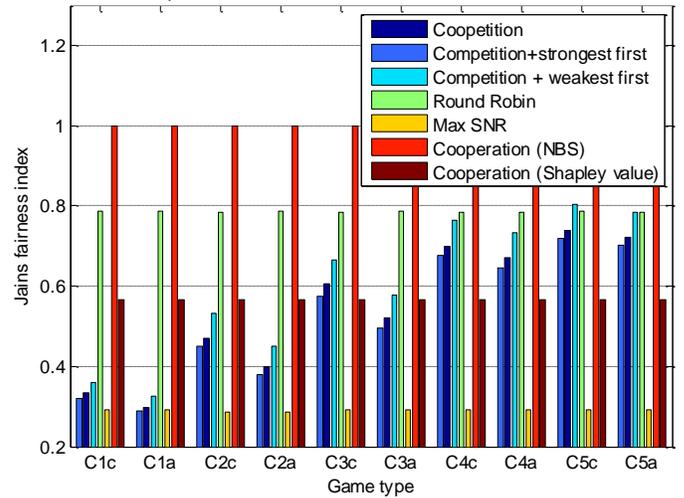

**Figure 11. Jain's fairness index vs. Cournot game type in coopetition Scenario 5.**

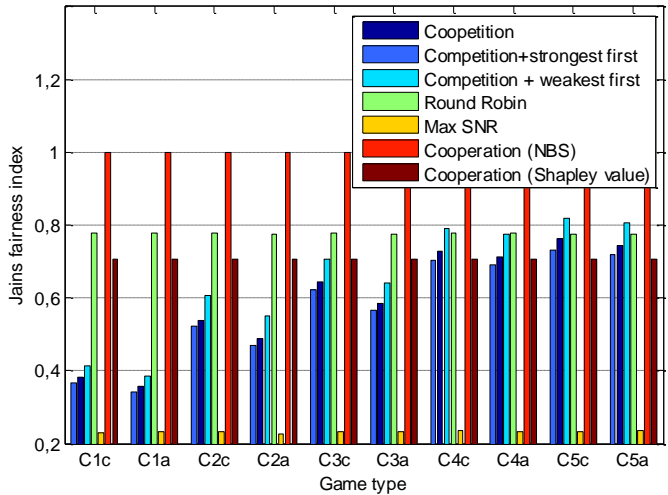

**Figure 9. Jain's fairness index vs. Cournot game type in coopetition Scenario 4.**

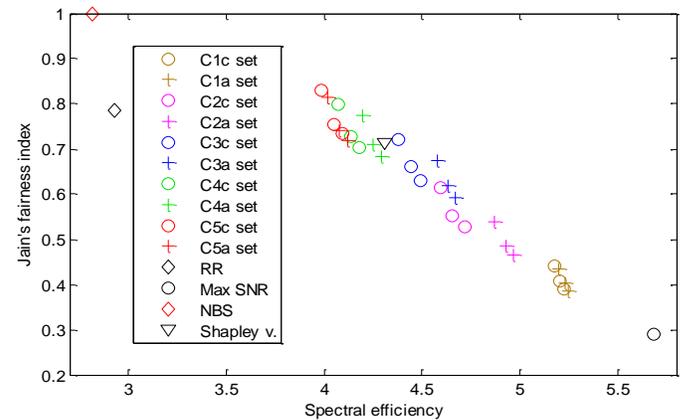

**Figure 12. Jain's fairness index vs. spectral efficiency in Scenario 3.**

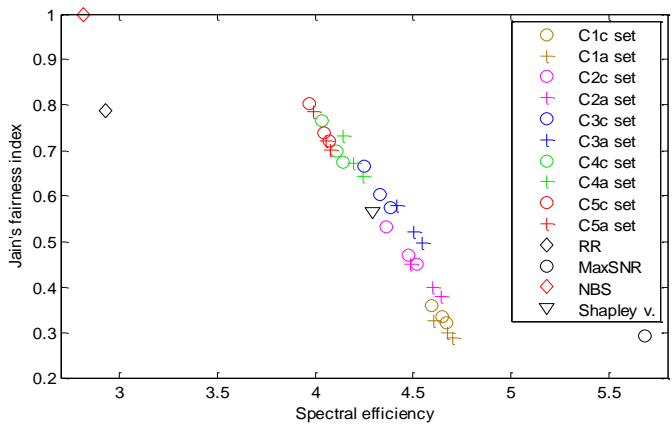

**Figure 13. Jain's fairness index vs. spectral efficiency in Scenario 5.**

## V. Conclusion

In this paper, we have presented a new coopetition methodology of spectrum sharing, which may be used in distributed CR networks. This practical methodology limits the transmission of the information required for efficient and fair allocation of resources, and allows the CR nodes to take decisions locally. The method has been tested for various network scenarios. It allows for balancing fairness and SE.

We believe that coopetition is a concept that can be successfully applied in many other scenarios and networks with different non-cooperative and cooperative game models. Future work will encompass optimization of the coopetiton parameters (e.g. $x$ and $y$ in the Cournot game or $\mu$ in the weighted-majority game) to maximize of the goal function in the network, e.g. sum-throughput or the SE or fairness. This problem is very complex and requires numerical optimization methods.